\documentclass[conference]{IEEEtran}
\IEEEoverridecommandlockouts

\usepackage{cite}
\usepackage{amsmath,amssymb,amsfonts}
\usepackage{algorithmic}
\usepackage{graphicx}
\usepackage{textcomp}
\usepackage{comment}
\usepackage{xcolor}
\usepackage{subfigure}
\usepackage{subcaption}
\usepackage{float}
\usepackage{fancyhdr}
\usepackage{multirow}
\usepackage{enumerate}
\usepackage{balance}
\usepackage[colorlinks, linkcolor=blue, citecolor=blue, urlcolor=blue]{hyperref}

\def\BibTeX{{\rm B\kern-.05em{\sc i\kern-.025em b}\kern-.08em
    T\kern-.1667em\lower.7ex\hbox{E}\kern-.125emX}}

\fancypagestyle{firstpage}
{
    \fancyhead[R]{} 
    \fancyhead[L]{2024 27th International Conference on Computer and Information Technology (ICCIT),\\ 20-22 December 2024, Cox’s Bazar, Bangladesh}
    \fancyfoot[C]{} 
    \fancyfoot[L]{979-8-3315-1909-4/24/\$31.00 ©2024 IEEE}
    \setlength{\footskip}{25pt} 
}

\begin{document}

\makeatletter
    \newcommand{\linebreakand}{%
      \end{@IEEEauthorhalign}
      \hfill\mbox{}\par
      \mbox{}\hfill\begin{@IEEEauthorhalign}
    }

 \makeatother
    
\title{From Images to Insights: Transforming Brain Cancer Diagnosis with Explainable AI}

\author{\IEEEauthorblockN{ Md. Arafat Alam Khandaker\IEEEauthorrefmark{1}, Ziyan Shirin Raha\IEEEauthorrefmark{1}, Salehin Bin Iqbal\IEEEauthorrefmark{1},  M.F. Mridha\IEEEauthorrefmark{2}, Jungpil Shin\IEEEauthorrefmark{3}
}
\IEEEauthorblockA{\IEEEauthorrefmark{1}Department of Computer Science and Engineering, Ahsanullah University of Science and Technology, Dhaka, Bangladesh}
\IEEEauthorblockA{\IEEEauthorrefmark{2}Department of Computer Science, American International University Bangladesh, Dhaka, Bangladesh}
\IEEEauthorblockA{\IEEEauthorrefmark{3}Department of Computer Science and Engineering, University of Aizu, Aizuwakamatsu, Japan}

aarafatalam18@gmail.com, ziyanraha@gmail.com, 
salehin.rawna@gmail.com, firoz.mridha@aiub.edu, jpshin@u-aizu.ac.jp
}

\maketitle
\thispagestyle{firstpage}
\begin{abstract}
Brain cancer represents a major challenge in medical diagnostics, requisite precise and timely detection for effective treatment. Diagnosis initially relies on the proficiency of radiologists, which can cause difficulties and threats when the expertise is sparse. Despite the use of imaging resources, brain cancer remains often difficult, time-consuming, and vulnerable to intraclass variability. This study conveys the Bangladesh Brain Cancer MRI Dataset, containing 6,056 MRI images organized into three categories: Brain Tumor, Brain Glioma, and Brain Menin. The dataset was collected from several hospitals in Bangladesh, providing a diverse and realistic sample for research. We implemented advanced deep learning models, and DenseNet169 achieved exceptional results, with accuracy, precision, recall, and F1-Score all reaching 0.9983. In addition, Explainable AI (XAI) methods including GradCAM, GradCAM++, ScoreCAM, and LayerCAM were employed to provide visual representations of the decision-making processes of the models. In the context of brain cancer, these techniques highlight DenseNet169's potential to enhance diagnostic accuracy while simultaneously offering transparency, facilitating early diagnosis and better patient outcomes.

\end{abstract}

\begin{IEEEkeywords}
Brain cancer, MRI Images, Medical Diagnostic, Explainable AI
\end{IEEEkeywords}
\section{Introduction}
\label{sec:introduction}
Brain cancer remains one of the most challenging and critical medical conditions having a wide variety of tumor types that poses a lot of challenges in medical diagnosis due to the complexity and the need for precise detection. Despite being advanced in imaging technology like Magnetic Resonance Imaging (MRI), the diagnosis of brain cancer remains challenging. The traditional methodology, which is the reliance on the radiologists’ expertise can sometimes be hindered by factors such as tumor diversity, limitations of human interpretation, particularly in regions with limited specialized resources \cite{1}. This scenario is even worsened by the need for timely, detailed and accurate classification of the cancer which is crucial for effective treatment planning and patient management \cite{2}.

Recent studies show that brain tumors, such as meningiomas, gliomas and other malignancies have significant intraclass variability making accurate identification challenging \cite{3}. As a result, there is an increasing demand for advanced computational tools to improve diagnosis accuracy and assist radiologists in their decision-making processes. Deep learning (DL) approaches have shown great potential for improving medical image analysis by automating tumor identification and classification \cite{4}. Convolutional Neural Networks (CNNs), a subset of DL, have been successfully applied to various imaging tasks, including brain tumor classification, by learning complex features from large datasets \cite{2}\cite{5}.

Our study is conducted on the Bangladesh Brain Cancer MRI Dataset\cite{6}, a comprehensive collection of 6,056 MRI images categorized into three distinct brain cancer types. As DL models are quite powerful, they frequently function as ``black boxes," which makes it challenging to understand how they make decisions. This may make it more difficult for users to accept and trust these models, particularly in crucial medical applications where knowing the reasoning behind a diagnosis is crucial \cite{7}\cite{8}. To address these challenges and enhance the reliability and transparency of our models, we employed Explainable AI (XAI) techniques \cite{9}, including GradCAM, GradCAM++, ScoreCAM, and LayerCAM. These methods provide visual explanations of the model's decision-making process, revealing how predictions are made and solving the `black box' problem. The research paper's contributions are summarized below: 
\begin{itemize}
 \item This study investigated nine pre-trained CNN models (DenseNet121, DenseNet169, DenseNet201, ResNet50, ResNet101, ResNet152, MobileNetV3, Xception, and InceptionV3) for the automated categorization of brain cancer in Brain MRI images.
 \item In order to boost interpretability, explainable AI approaches, such as Grad-CAM, Grad-CAM++, Score-CAM, and Layer CAM have been implemented.
 \item Our research aims to address the gap in explainable AI methods for the classification of brain cancer in Brain MRI images.
 \end{itemize}

 This study begins with an introduction in Section \ref{sec:introduction}, followed by a review of related works in Section \ref{sec:Related Works}. The  background study is detailed in Section \ref{sec:BackGround Study}, methodology is detailed in Section \ref{sec:methodology}, results are analyzed in Section \ref{sec:result analysis}, and limitations with future work are discussed in Section \ref{sec:limitation}. Finally, Section \ref{sec:conclusion} summarizes the study.

\section{Related Works}
\label{sec:Related Works}
This section gives  a compact overview of previous study on Brain Tumor Classification which is relevant to our study.
\subsection{ Brain Tumor Classification without using XAI Techniques 
}
Vinod Kumar et al. \cite{4} , Saleh et al. \cite{10} , Heba et al. \cite{11}, and Sneha et al. \cite{13}  each proposed methods for detecting brain tumors using various deep learning models \cite{12}, without integrating Explainable AI (XAI) techniques. Vinod Kumar et al. applied transfer learning using CNNs such as AlexNet, VGG-16, and ResNet-50, and introduced a hybrid model combining VGG-16 and ResNet-50. Their approach, validated on a Kaggle dataset containing 3,264 MRI images, achieved impressive accuracy, sensitivity, and specificity of 99.98\%. Saleh et al. explored the classification of brain tumors by leveraging pre-trained CNNs- Xception, ResNet50, InceptionV3, VGG16, and MobileNet on a Kaggle dataset of 4,480 MRI images, resulting in F1-scores between 97.25\% and 98.75\%. Heba et al. suggested a technique that combines a Deep Neural Network (DNN) with Discrete Wavelet Transform (DWT) for feature extraction and uses Principal Component Analysis (PCA) for feature reduction. This method, when tested on 66 MRI images from Harvard Medical School, yields a classification rate of 96.97\%. Lastly, Sneha et al. compared a basic CNN with VGG-16 architecture for brain tumor detection, training on an augmented dataset of 2,065 MRI images. Their basic CNN achieved 93.36\% training accuracy and 86.45\% validation accuracy, while VGG-16 reached 97.16\% training accuracy and 97.42\% validation accuracy.

\subsection{Brain Tumor Classification using XAI Techniques }
Fahad et al. \cite{8} , Ramy A et al. \cite{14} , Eman Ragab et al. \cite{2}  and Burak et al. \cite{16} each proposed methods for detecting brain tumors using various deep learning models incorporating XAI techniques. Fahad et al. used the VGG-16 model with transfer learning and Layer-wise Relevance Propagation (LRP) to detect brain tumors from 3,000 MRI images, achieving 97.33\% accuracy. Ramy A et al. introduced the NeuroXAI framework, which integrated seven XAI techniques, such as GradCAM and SmoothGrad, with CNN models like ResNet-50, to enhance the interpretability of brain tumor MRI analysis. Eman Ragab et al. utilized CNNs, including VGG16, VGG19, and ResNet50, to classify pediatric posterior fossa tumors from a dataset of 300,000 MRI images. VGG16 reached the highest accuracy of 95.33\%. Burak et al. presented DGXAINet, which combined DenseNet201 for feature extraction, GradCAM for visualization, and Iterative Neighborhood Component Analysis (INCA) for feature selection, achieving accuracies of 98.65\% and 99.97\% on two different datasets. Collectively, these studies illustrate how integrating deep learning with XAI can enhance both classification performance and model transparency in brain tumor detection.

\setcounter{figure}{1}
\begin{figure*}[htbp]
\centering
\includegraphics[width=1\textwidth]{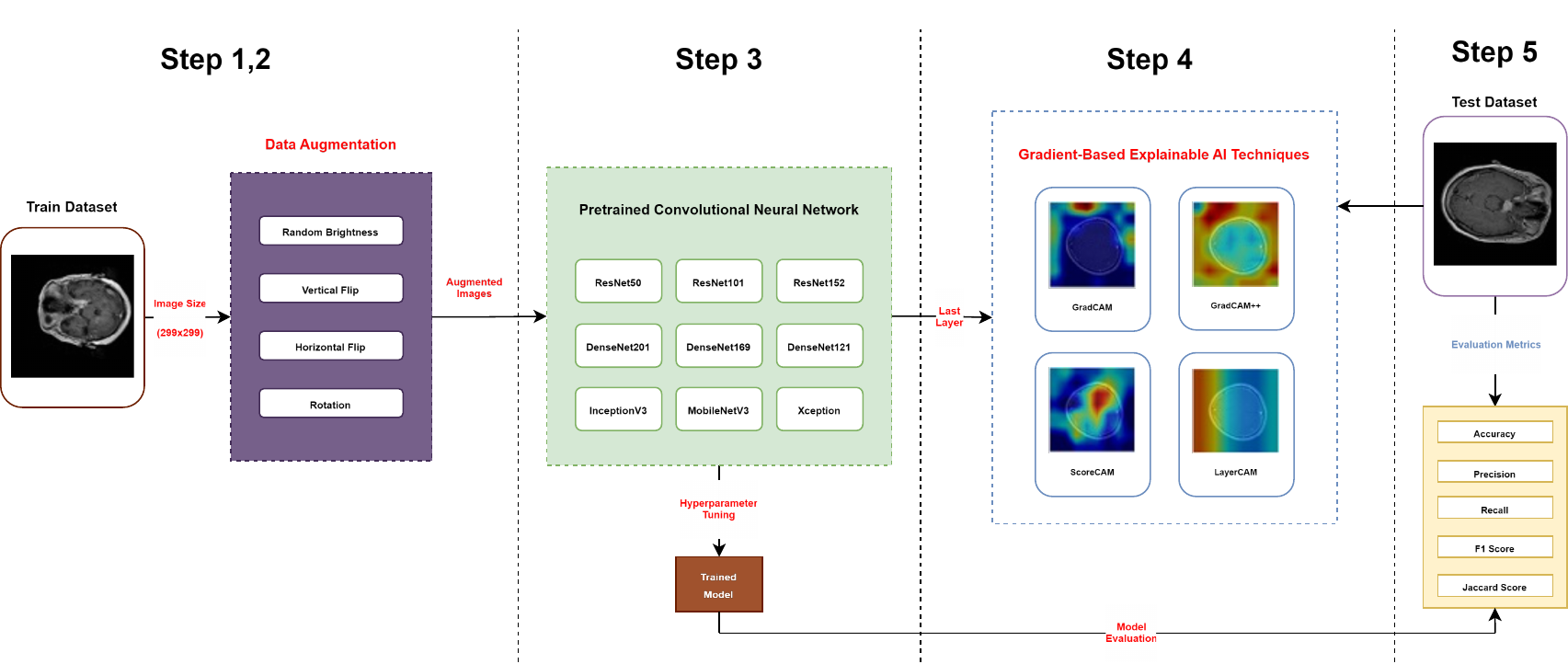}
\caption{Methodological Framework for Brain Cancer Classification}
\label{fig:methodology}
\end{figure*}

\section{BackGround Study }
\label{sec:BackGround Study}
\subsection{Convolutional Neural Networks (CNNs)}
Numerous CNN-based methods for brain cancer detection, including DenseNet \cite{17}, ResNet \cite{18}, MobileNet \cite{19}, Inception \cite{20}, and Xception \cite{21}, have achieved significant accuracy improvements. ResNet models (e.g., ResNet50, ResNet101, ResNet152) address the vanishing gradient issue using residual connections, enabling deeper networks. DenseNet models (e.g., DenseNet121, DenseNet169, DenseNet201) enhance efficiency through cross-layer feature reuse. MobileNetV3 reduces model size and computational complexity for resource-limited settings. InceptionV3 and Xception leverage multi-scale feature extraction and depth-wise separable convolutions to balance accuracy and efficiency. Hybrid CNN models further advance brain cancer detection \cite{15}.
\subsection{Explainable Artificial Intelligence (XAI) techniques}

Explainable AI (XAI) techniques enhance the interpretability of deep learning models in medical imaging, particularly for brain cancer detection. Methods like GradCAM, GradCAM++, ScoreCAM, and LayerCAM visualize critical MRI regions used in model predictions \cite{22}. GradCAM generates heatmaps \cite{23} highlighting essential areas, while GradCAM++ improves localization by considering both positive and negative pixel influences \cite{24}. ScoreCAM assigns importance scores to spatial regions, further refining interpretability. LayerCAM provides deeper insights by assigning relevance scores across network layers, clarifying how image features impact predictions.
\subsection{Evaluation Metrics}
Accurate evaluation of brain cancer classification models depend on several essential  metrics. Accuracy measures the proportion of correct predictions out of all cases, showing the model's overall effectiveness. Precision and recall assess the model's ability to identify brain tumors accurately, with precision focusing on reducing false positives and recall on reducing false negatives. The F1 score, which combines precision and recall, is especially valuable when dealing with imbalanced classes. Additionally, the Jaccard score measures how well predicted tumor classifications overlap with actual results, making it specially useful for multi-class and imbalanced datasets.

\section{Methodology }
\label{sec:methodology}
\subsection{Dataset}

``The Bangladesh Brain Cancer MRI Dataset'' contains 6,056 MRI pictures divided into three categories: Brain Glioma (2,004 images), Brain Menin (2,004 images), and Brain Tumor (2,048 images). This dataset, assembled from various hospitals in Bangladesh, consistently downsized to 512x512 pixels, provides a broad and accurate representation for brain cancer research. It was created in alliance with medical experts to ensure accuracy and usefulness, making it an valuable resource for creating diagnostic tools in medical imaging.

\setcounter{figure}{0}
\begin{figure}[H]
    \centering
    \subfigure[Brain Menin]{
        \includegraphics[width=0.11\textwidth]{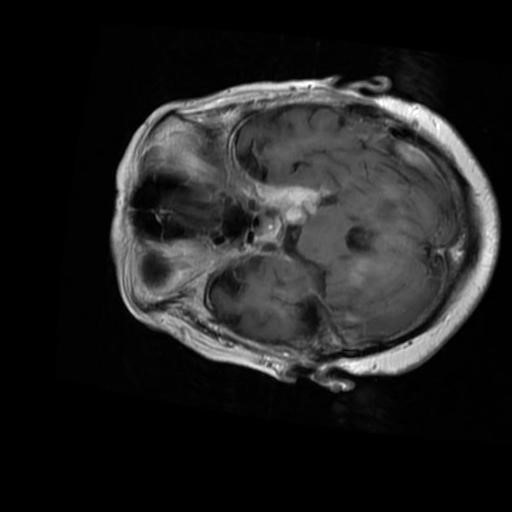}
        \label{fig:sub1}
    }
    \subfigure[Brain Glioma]{
        \includegraphics[width=0.11\textwidth]{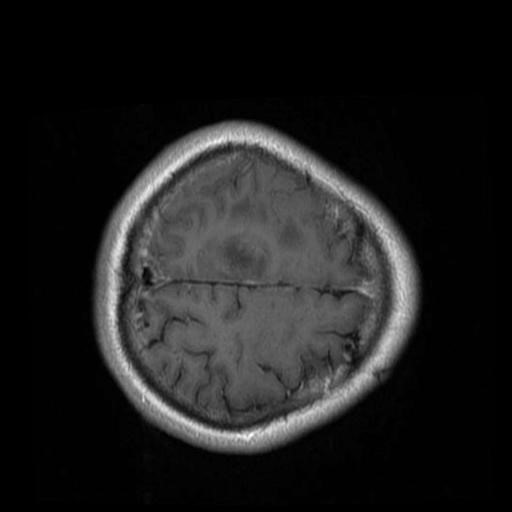}
        \label{fig:sub2}
    }
    \subfigure[Brain Tumor]{
        \includegraphics[width=0.11\textwidth]{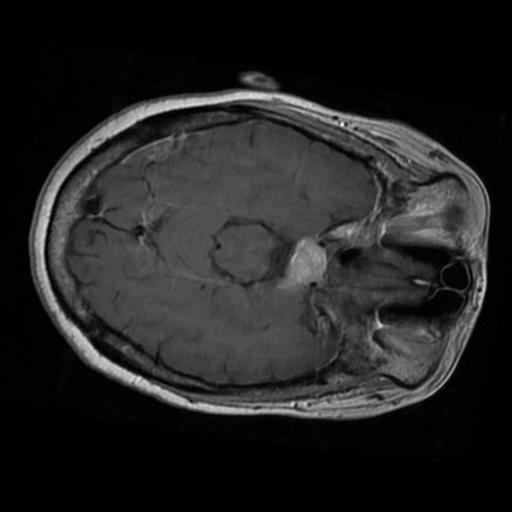}
        \label{fig:sub3}
    }
    \caption{Dataset Representation}
    \label{fig:dataset_example}
\end{figure}

\subsection{ Proposed Methodology}
To effectively compare the performance of various models in analyzing brain cancer images, we designed a methodology, as defined in Fig. \ref{fig:methodology} of our study.

\textbf{Step 1)  Preparing Input Image:}
For keeping stability of the different models and approaches, initial processing processes were applied to every image, starting with adjusting to a standard 256 x 256 pixel size. 

\textbf{Step 2) Image Preprocessing Technique:} Normalizing pixel values 0 to 1 scale assisted to guarantee a steady flow of data and increase the rate of closure during training.

\textbf{Data Augmentation:} The model's capacity to generalize to various orientations was enhanced by applying random rotations to add diversity in image orientation. In addition, mirrored images were simulated via horizontal flipping, which increased dataset variety and reduced the likelihood of overfitting.

\textbf{Cropping and Focus Enhancement:} Images were selectively cropped to remove additional background elements and enhance areas of interest. This allowed us to focus on key features while reducing computational overhead.

\textbf{Brightness Adjustment:} Image brightness adjustments were applied to limit the impact of lighting fluctuations across the dataset, resulting in uniform exposure levels.

\textbf{Contrast Enhancement:} These techniques were used to improve image contrast, which is important in medical imaging since it allows for precise classification of minor characteristics.

\textbf{Step 3) Model Selection and Training:} For this study we opted nine  Convolutional Neural Network (CNN) architectures known for their proficiency in image classification tasks: Xception, DenseNet201, DenseNet121, DenseNet169, MobileNetV3, InceptionV3, ResNet50, ResNet101 and ResNet152.\newline The preprocessed dataset was used throughout training to ensure stability and consistency. Hyperparameter optimization was used to fine-tune each model for the specific objective of classifying brain cancers. 

\textbf{Step 4) Application of Explainable AI (XAI) Techniques:} To improve the understanding of the model’s decisions, we employed various Explainable AI techniques on the final layers of the CNNs. GradCAM, GradCAM++, ScoreCAM, and LayerCAM were used to generate heatmaps that emphasize the most critical portions of the images for each class, allowing us to see which elements of the image influenced the model's decision-making the most. These strategies provided class-specific attention mapping and multi-layer activation visualization, allowing for more in-depth insights into model performance.

\textbf{Step 5) Performance Evaluation:} We used a variety of performance indicators to assess each model's ability to recognize brain cancer photos. This detailed review enabled us to analyze and compare the performance of many models, allowing us to choose the best one for our categorization task.\\

\section{Result Analysis}
\label{sec:result analysis}

\subsection{Experimental setup}

Experiments carried out on Kaggle Notebooks platform with Tesla P100 GPU computing processor. The dataset is divided into train, validation and test instances. 80\% of the dataset is divided into training set while the remaining 20\% of the dataset is divided in half between the validation set (10\%) and test set (10\%).

The suggested models classifies brain MRI images into brain tumor, brain glioma and brain menin. The training parameters for the suggested models, including the number of epochs, optimization algorithm, input image size, batch size, and learning rate, are detailed in TABLE \ref{tab1}.


\begin{table}[htbp]
\caption{Hyperparameter Settings of Different Models}
\begin{center}
\setlength{\tabcolsep}{3.5pt}
\renewcommand{\arraystretch}{1.2}
\begin{tabular}{|c|c|c|c|}
\hline
\textbf{Models} & \textbf{Learning Rate} & \textbf{Number of Epochs} & \textbf{Optimizer} \\
\hline
DenseNet201 & 0.0001 & 15 & Adam \\
\hline
DenseNet169 & 0.0001 & 10 & Adam \\
\hline
ResNet152 & 0.0001 & 11 & Adam \\
\hline
ResNet101 & 0.0001 & 15 & Adam \\
\hline
MobileNetV3 & 0.0001 & 14 & Adam \\
\hline
InceptionV3 & 0.0001 & 15 & Adam \\
\hline
ResNet50 & 0.0001 & 12 & Adam \\
\hline
Xception & 0.0001 & 13 & Adam \\
\hline
DenseNet121 & 0.0001 & 15 & Adam \\
\hline
\end{tabular}
\label{tab1}
\end{center}
\end{table}

\subsection{Evaluation}

The evaluation is conducted by considering performance metrics such as accuracy, precision, recall, F1-score and jaccard score which are presented in TABLE \ref{tab2} and juxtaposed in Fig. \ref{fig:performance_comparison} for comparative purposes.

\begin{table}[htbp]
\caption{Performance Evaluation of Different Models}
\begin{center}
\setlength{\tabcolsep}{3.5pt}
\renewcommand{\arraystretch}{2.2}
\begin{tabular}{|c|c|c|c|c|c|}
\hline
\textbf{Architecture} & \textbf{Accuracy} & \textbf{Precision} & \textbf{Recall} & \textbf{F1-Score} & \textbf{Jaccard Score} \\
\hline
DenseNet201 & 0.9934 & 0.9934 & 0.9934 & 0.9934 & 0.9869 \\
\hline
\textbf{DenseNet169} & \textbf{0.9983} & \textbf{0.9983} & \textbf{0.9983} & \textbf{0.9983} & \textbf{0.9967} \\
\hline
ResNet152 & 0.9851 & 0.9853 & 0.9851 & 0.9851 & 0.9707 \\
\hline
ResNet101 & 0.9901 & 0.9902 & 0.9901 & 0.9901 & 0.9804 \\
\hline
MobileNetV3 & 0.9258 & 0.9298 & 0.9258 & 0.9244 & 0.8605 \\
\hline
InceptionV3 & 0.9884 & 0.9887 & 0.9884 & 0.9884 & 0.9772 \\
\hline
ResNet50 & 0.9917 & 0.9917 & 0.9917 & 0.9917 & 0.9836 \\
\hline
Xception & 0.9950 & 0.9950 & 0.9950 & 0.9950 & 0.9901 \\
\hline
DenseNet121 & 0.9917 & 0.9917 & 0.9917 & 0.9917 & 0.9836 \\
\hline
\end{tabular}
\label{tab2}
\end{center}
\end{table}

\setcounter{figure}{2}
\begin{figure}[htbp]
\setlength{\fboxsep}{0pt}
\setlength{\fboxrule}{1pt}
\centering
\fcolorbox{lightgray}{white}{\includegraphics[width=0.485\textwidth]{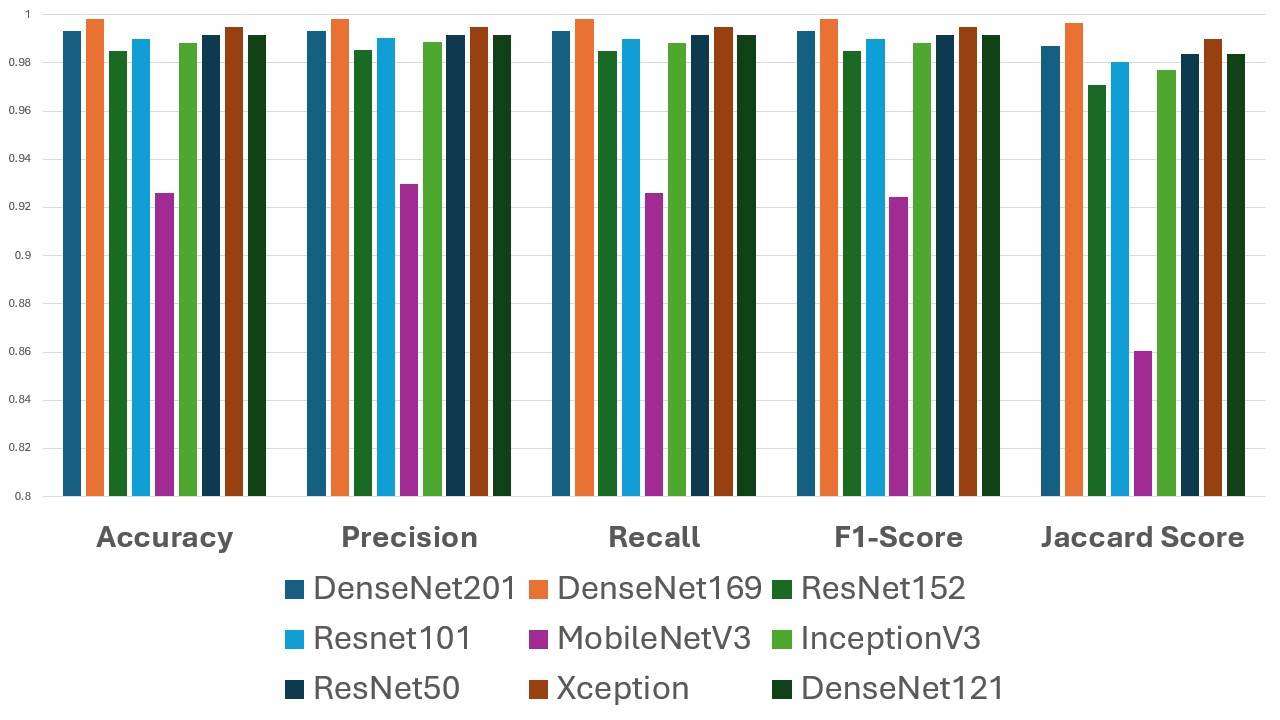}}
\caption{Performance Analysis Graph of the Proposed Models}
\label{fig:performance_comparison}
\end{figure}

Based on the analysis, the DenseNet169 model demonstrated remarkable performance compared to the other models. Specifically, it achieved an accuracy of 0.9983 and F1 score of 0.9983, indicating its ability of precisely classifying the instances. Moreover, the model attained a jaccard score of 0.9967, which highlights its capability of correctly identifying the similarity and diversity of sample sets. In contrast, the following models - DenseNet201, ResNet152, ResNet101, InceptionV3, ResNet50, Xception, DenseNet121 - also performed well, though they are slightly behind of DenseNet169 in terms of performance metrics. Xception, which achieved an accuracy of 0.9950, an F1-score of 0.9950 and a jaccard score of 0.9901 is the second best performing model of our study. DenseNet201, ResNet152, ResNet101, InceptionV3, ResNet50 and DenseNet121 also shows competitive results, all achieving high scores across all metrics, particularly with accuracy ranging from 0.9851 to 0.9934 and F1-scores between 0.9851 and 0.9934. This indicates that these models can perform well in brain cancer classification but do not surpass the performance of DenseNet169. On the other hand, the MobileNetV3 model having an accuracy of 0.9258 and F1 score of 0.9244, has shown the least favorable performance of all the models. These numbers suggest that MobileNetV3 is less effective for this particular classification task compared to the other models. Fig. \ref{fig:confusionmatrix} shows the testing confusion matrix on different architectures of the proposed model. DenseNet169 emerges because of its densely connected layers, efficient feature reuse, low parameter redundancy, and ability to capture complicated spatial features, making it perfect for high-resolution image classification.

\begin{figure*}[htbp]
    \centering
    \subfigure[DenseNet201]{
        \fcolorbox{lightgray}{white}{\includegraphics[width=0.25\textwidth]{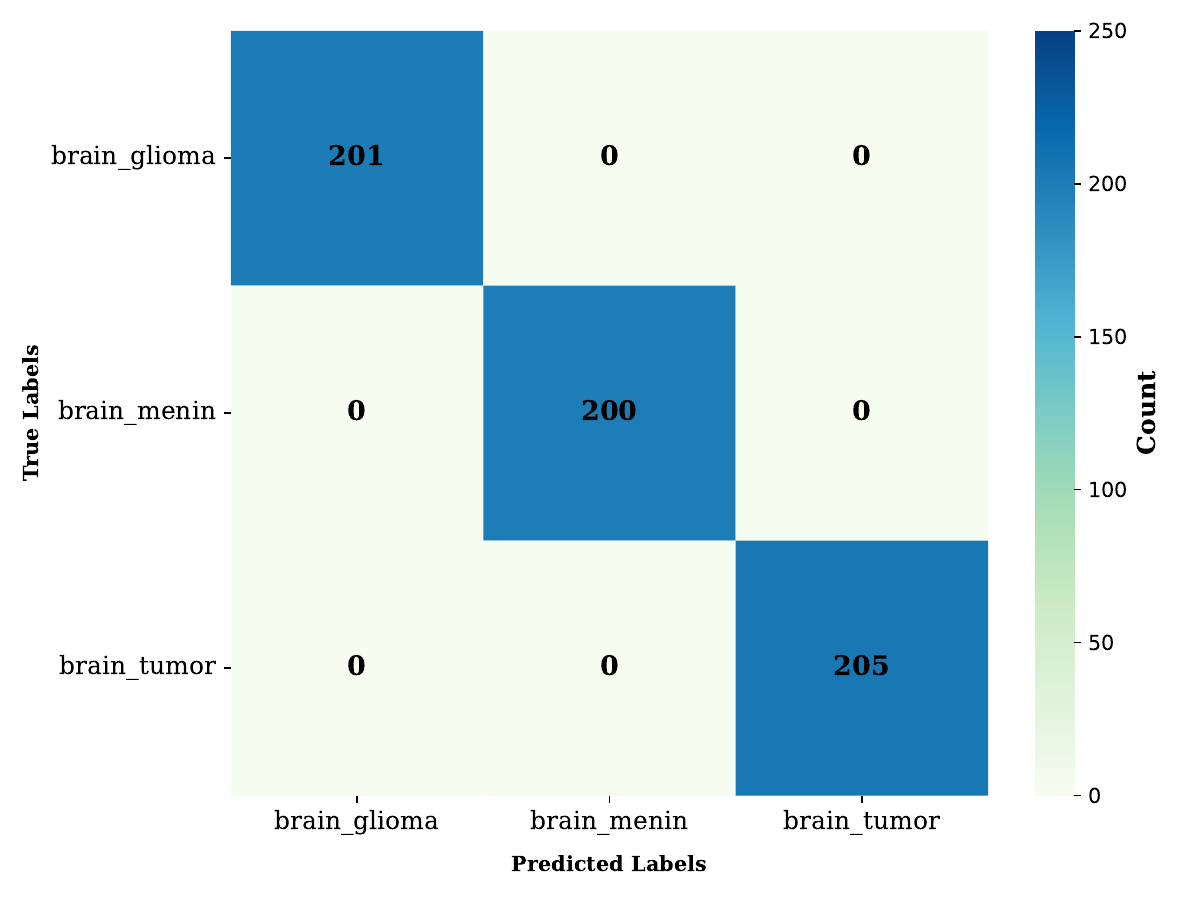}}
        \label{fig:sub1}
    }
    \subfigure[DenseNet169]{
        \fcolorbox{lightgray}{white}{\includegraphics[width=0.24\textwidth]{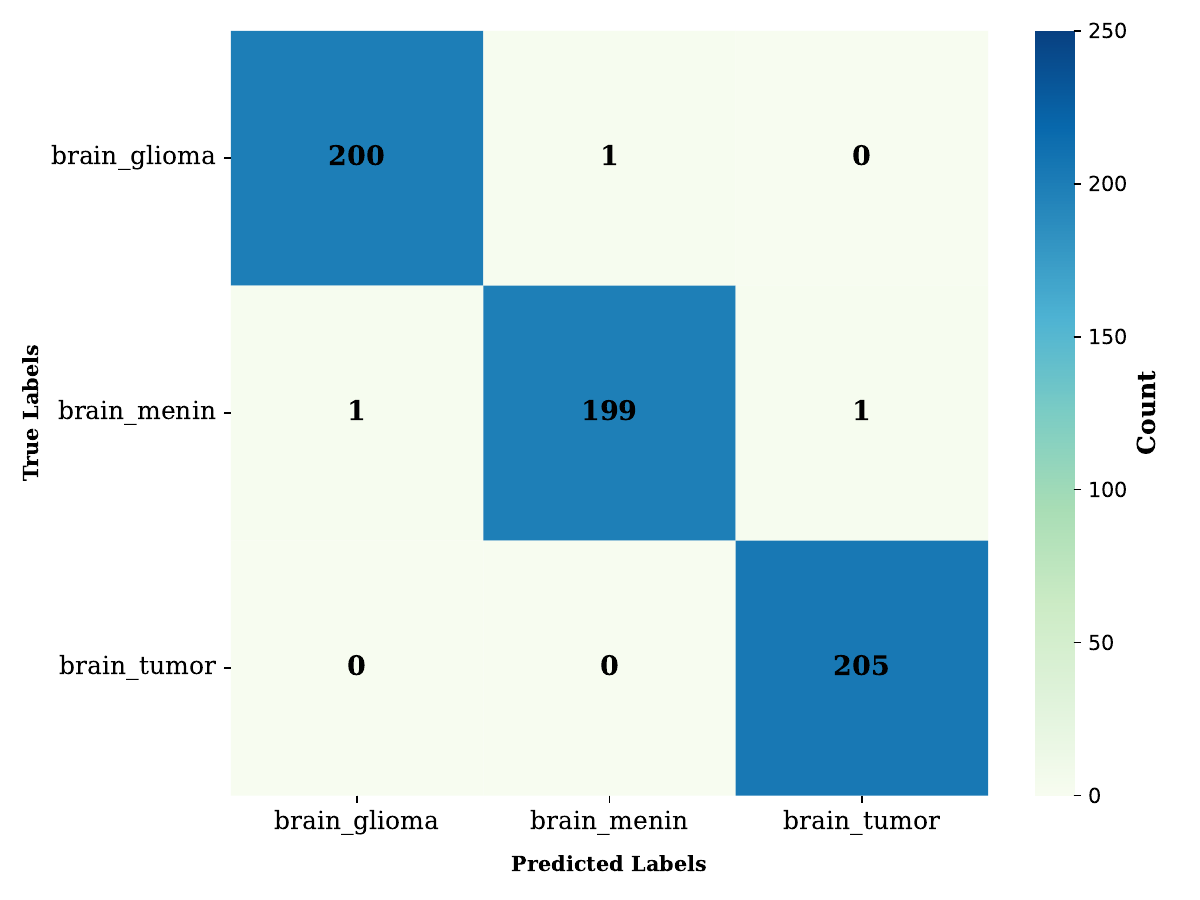}}
        \label{fig:sub2}
    }
    \subfigure[ResNet152]{
        \fcolorbox{lightgray}{white}{\includegraphics[width=0.24\textwidth]{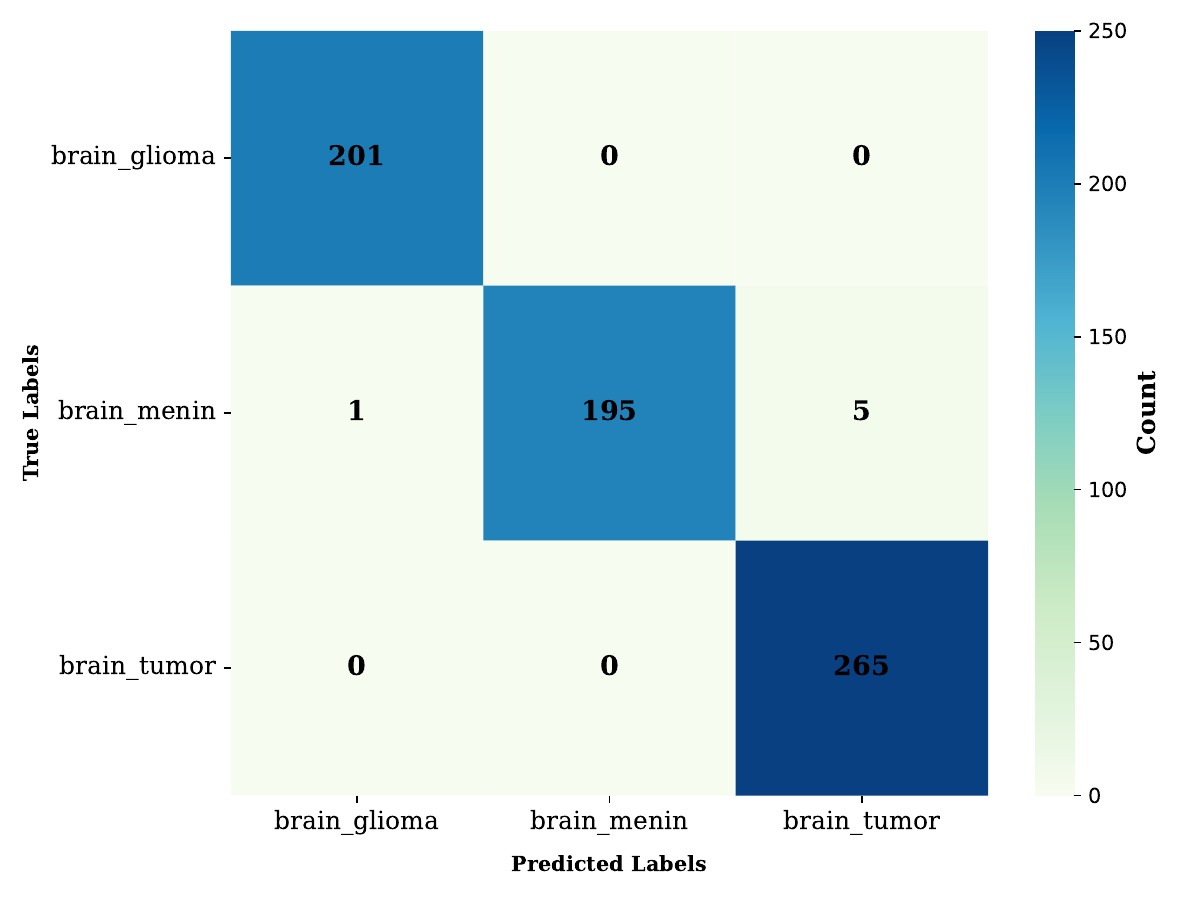}}
        \label{fig:sub3}
    }
    \subfigure[ResNet101]{
        \fcolorbox{lightgray}{white}{\includegraphics[width=0.24\textwidth]{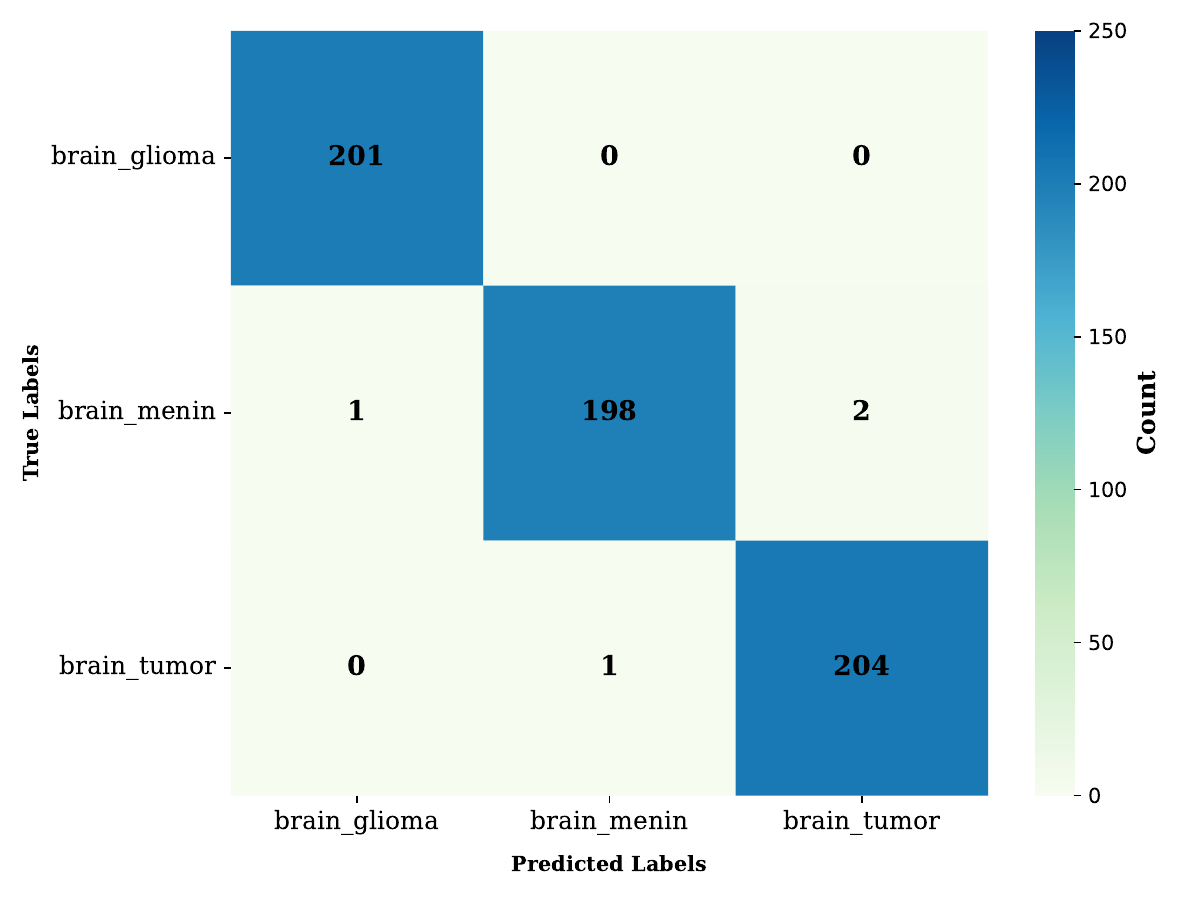}}
        \label{fig:sub4}
    }
    \subfigure[MobileNetV3]{
        \fcolorbox{lightgray}{white}{\includegraphics[width=0.24\textwidth]{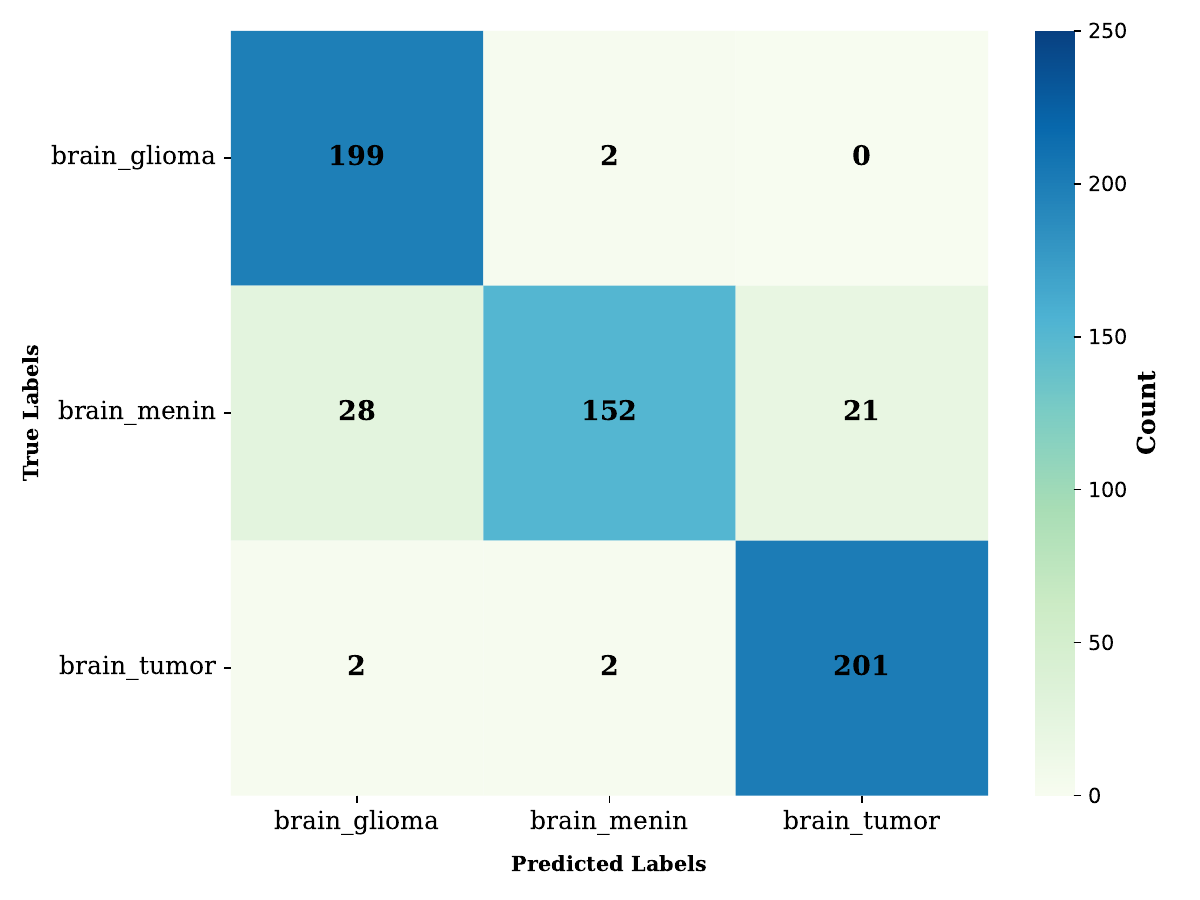}}
        \label{fig:sub5}
    }
    \subfigure[InceptionV3]{
        \fcolorbox{lightgray}{white}{\includegraphics[width=0.24\textwidth]{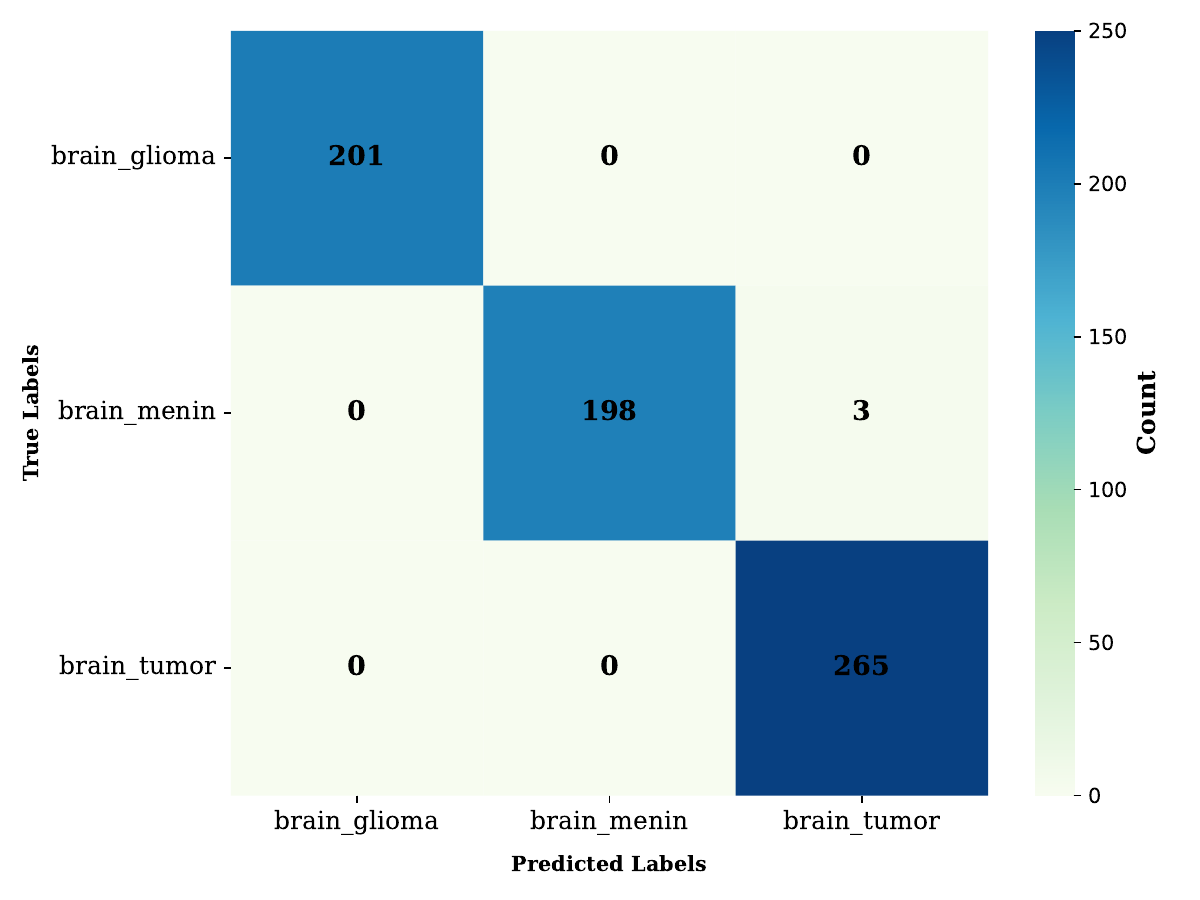}}
        \label{fig:sub6}
    }
    \subfigure[ResNet50]{
        \fcolorbox{lightgray}{white}{\includegraphics[width=0.24\textwidth]{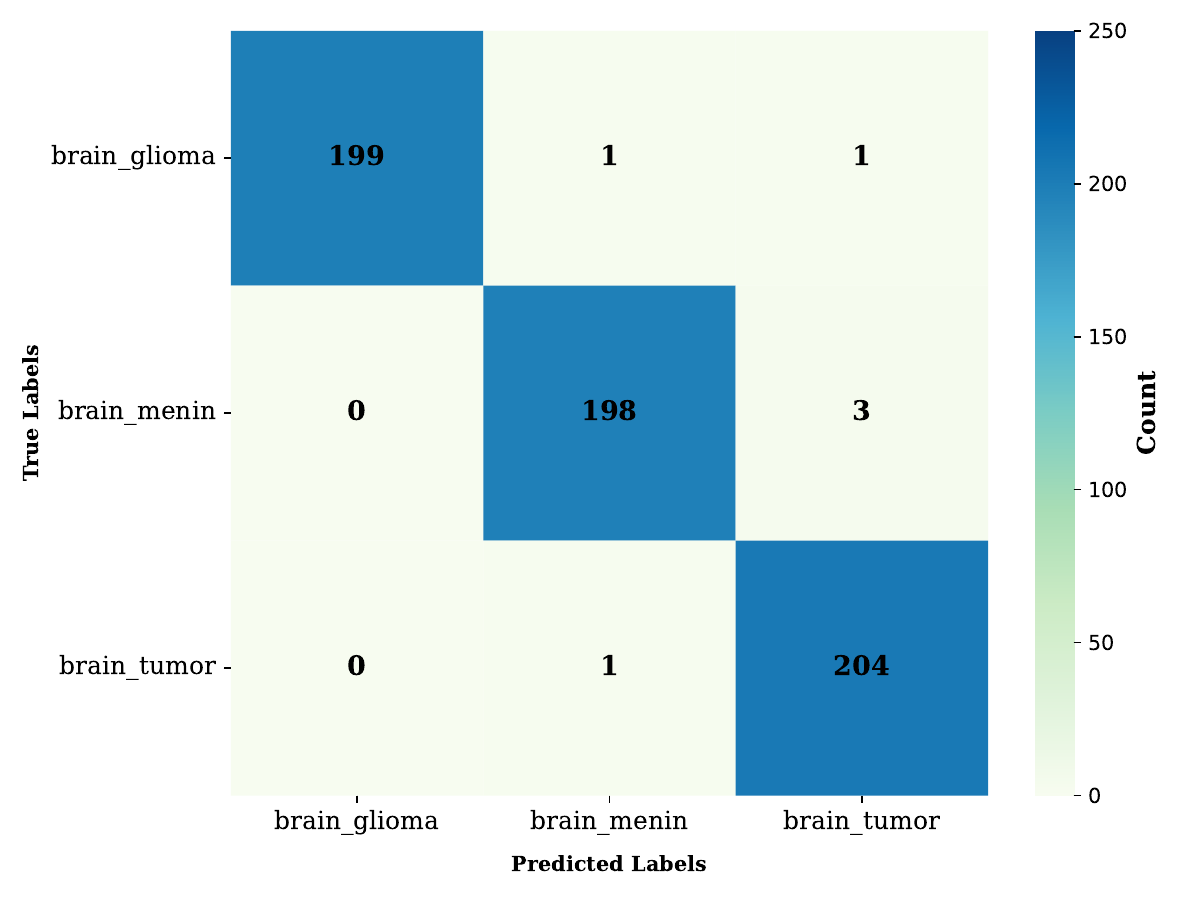}}
        \label{fig:sub7}
    }
    \subfigure[Xception]{
        \fcolorbox{lightgray}{white}{\includegraphics[width=0.24\textwidth]{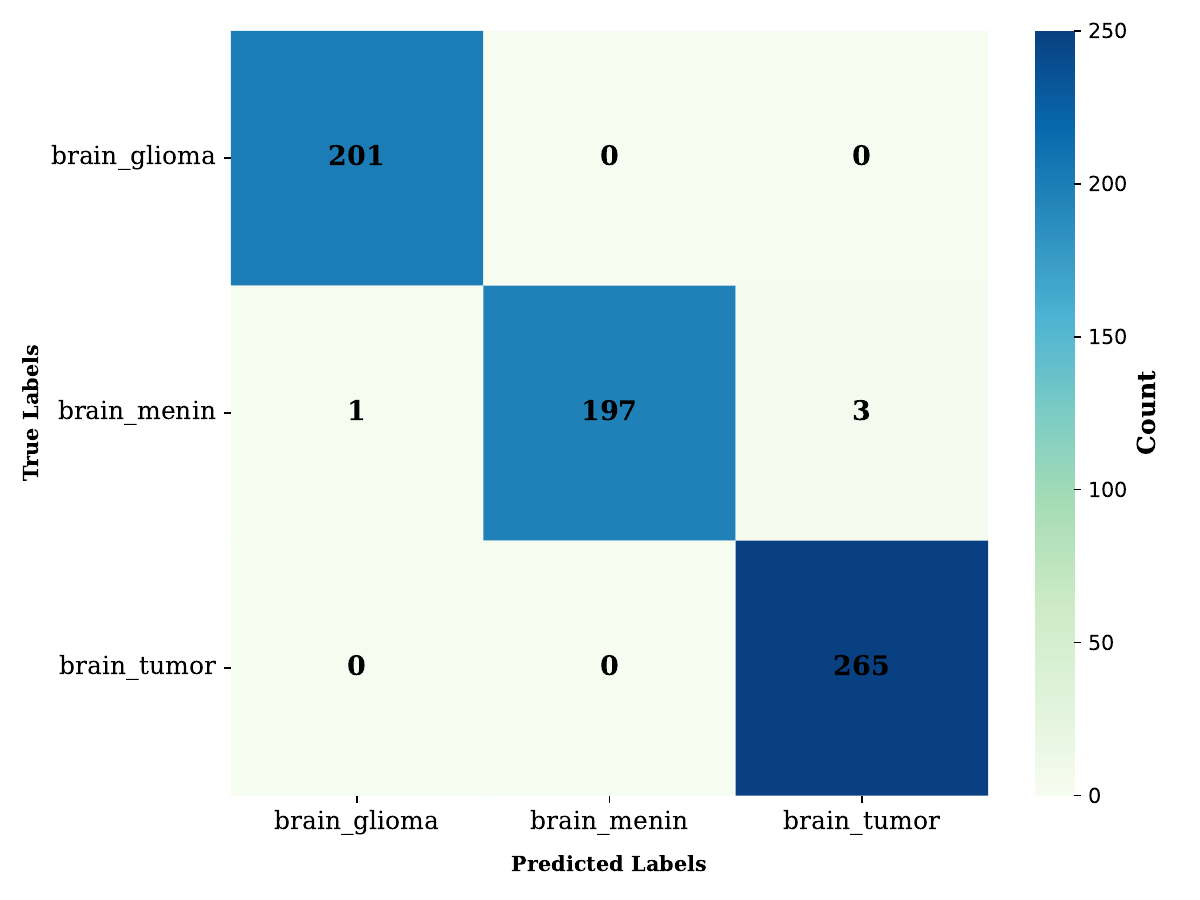}}
        \label{fig:sub8}
    }
    \subfigure[DenseNet121]{
        \fcolorbox{lightgray}{white}{\includegraphics[width=0.24\textwidth]{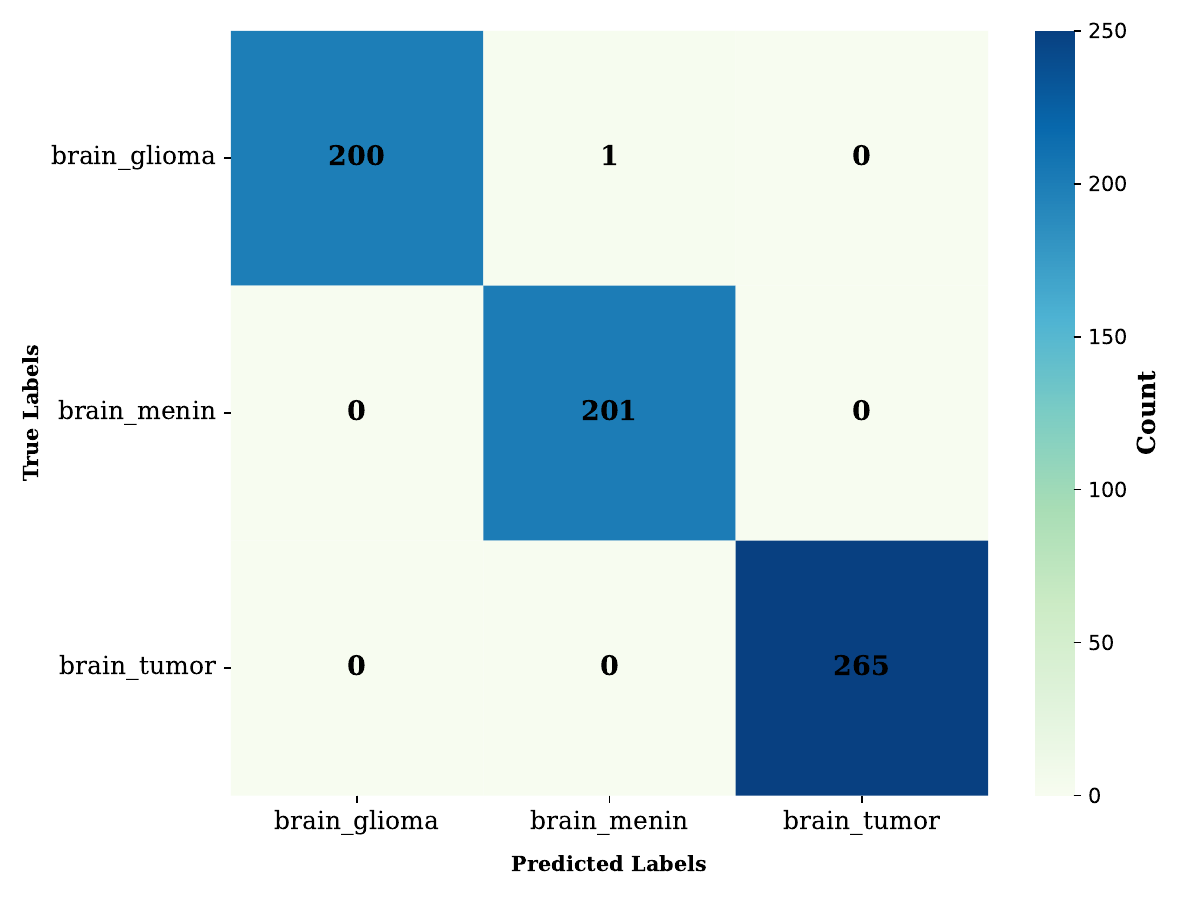}}
        \label{fig:sub9}
    }
    \caption{Confusion Matrix of the Proposed Models}
    \label{fig:confusionmatrix}
\end{figure*}

\subsection{Model Interpretation using XAI}
The usage of XAI technique, including GradCAM, GradCAM++, LayerCAM and ScoreCAM provides valuable insights into the decision making of the models. These tactics enhanced the visualization of distinct areas of the images, helping the detection of the  brain cancer more efficient and effective. The model's decision-making process is shown in Fig. \ref{fig:explanation_using_XAI}. These visualizations illustrate the model's decision-making process by emphasizing the regions of brain cancer images that have a significant effect on the classification results. The interpretability of the model is improved by these visual explanations, which emphasize the patterns that are most suggestive of specific malignant regions.

\begin{figure}[H]
    \centering
    \subfigure[GradCAM]{
        \includegraphics[width=0.15\textwidth]{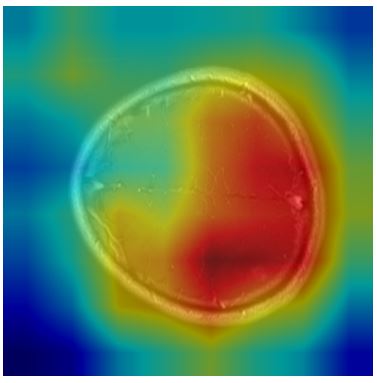}
        \label{fig:sub1}
    }
    \subfigure[GradCAM++]{
        \includegraphics[width=0.15\textwidth]{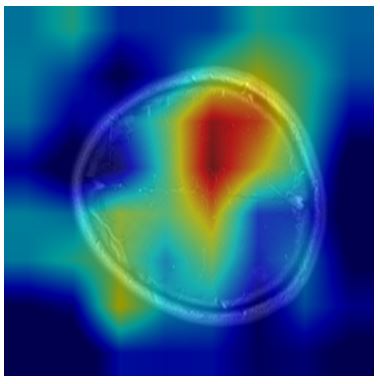}
        \label{fig:sub2}
    }
    \\
    \subfigure[LayerCAM]{
        \includegraphics[width=0.15\textwidth]{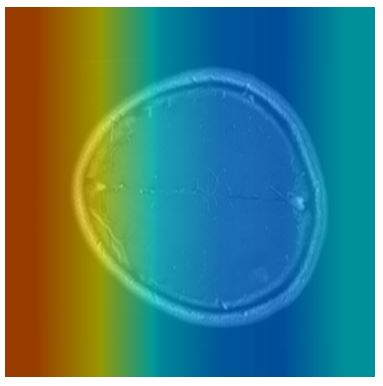}
        \label{fig:sub3}
    }
    \subfigure[ScoreCAM]{
        \includegraphics[width=0.15\textwidth]{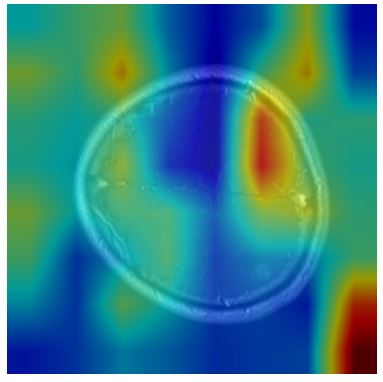}
        \label{fig:sub4}
    }
    \caption{Explanation of the Model's Decision Making}
    \label{fig:explanation_using_XAI}
\end{figure}

\section{Limitations \& Future Work}
\label{sec:limitation}
Despite of achieving an excellent result in classifying brain cancer from MRI images, our study faces several limitations. Firstly, the dataset's size and diversity is limited which could be affecting the model's generalization to broader populations. Expanding the dataset by incorporating more MRI images from additional locations and healthcare organizations, would enhance the robustness and adaptability. Incorporating a neutral category, such as MRI images of normal brains, may also improve the model's capacity to distinguish between pathological and healthy brain states, hence increasing classification accuracy. Moreover, the study focused on brain cancers, specifically meningioma, glioma and tumor classification. Extending our methods to other cancer kinds and illnesses would broaden the scope of the study.

Secondly, our current approach is limited to MRI image classification, ignoring other significant data sources such as genetic markers, histological data or patient's clinical histories. Future study could benefit from the incorporation of multimodal data, resulting in a more comprehensive diagnosis framework with higher classification accuracy.

Thirdly, while we used several Explainable AI (XAI) methods such as GradCAM, GradCAM++, ScoreCAM and LayerCAM to improve model transparency, interpretability issues remain. Medical judgments based exclusively on model outputs demand a higher level of clarity and dependability. Future research can also focus on enhanced XAI algorithms that provide more interpretable and clinically relevant reasons for model predictions.

Moving forward, exploration of novel deep learning architectures, expanding the dataset with normal brain images or other significant data, implementing new XAI techniques, these improvements will be crucial for translating our work into practical healthcare applications.

\section{Conclusion }
\label{sec:conclusion}
Our study presents a robust approach for classifying brain cancer using advanced deep learning and Explainable AI (XAI) techniques to enhance both accuracy and interpretability. Among the models tested, DenseNet169 achieved the highest accuracy of 0.9983, while MobileNetV3 had the lowest accuracy at 0.9258. Other models included DenseNet201 with 0.9934, DenseNet121  at 0.9917, Xception at 0.9950, ResNet152 at 0.9851, ResNet101 at 0.9901, InceptionV3 at 0.9884 and ResNet50 at 0.9917. We utilized XAI techniques such as GradCAM, GradCAM++, ScoreCAM and LayerCAM to clarify  how the CNN models make their predictions. These methods  enhanced  the visualization of important characteristics in microscopic images, helping to differentiate between malignant and benign tissues. This enhanced clarity offers improved  healthcare treatment. Overall, our study underscores the usefulness  of advanced deep learning and XAI in medical image analysis, highlighting superior performance and the advantages of XAI for improved diagnostic accuracy and clinical support.

\balance

\vspace{12pt}

\end{document}